# The Idea of Exo-brain

**Ankur Betageri**
Bharati College, University of Delhi, India

2022

# Abstract

In this paper I examine the process of getting affected by and the process of making sense of non-language sounds and propose the idea of the contextual cognitive apparatus or exo-brain. The affective power of the song comes from an affective reaction between the singing voice, especially the sung non-language sounds, and the linguistic and cultural contexts of the singer and the listener and out this affective reaction something emerges which constitutes the meaning, sensible or affective, of the non-language sound. I show how the cognitive contexts of the production and the reception of the non-language sounds, and the song, play a central role in our apprehension of the sounds and propose the existence of an extra-bodily sense making organ of our cognitive system, the extension of our brain and sense organs, which I call contextual cognitive apparatus or exo-brain.

**Keywords**: Sense-making, Intensification of affect, Non-language sound perception, Semantic localization, Contextual cognitive apparatus, Exo-brain.

## § 1. The Problem of Incongruity between Sung/Heard and Written Forms of the Song which Makes Interpretation Necessary

We do not know whether it is the work of the creative mouth or the creative ear but what we hear in the sung song is often incongruous with the song as it is written. There is greater affective freedom, semantic flexibility and musical non-comprehension in what can be sung and misheard, or unheard as language, and this musical and affective freedom is conquered by the singing voice. According to convention what gets written is what gets read so when one is armed with the singing voice and the spoken word one knows that there is greater semantic flexibility, musical non-comprehension and inflectional possibilities in parole than in the assured and assuring conventionality of langue.

Often it is only the singer's voice which gets represented in the written text of the song, and which constitutes the written existence of the song as such. But when the song exists just as authentically in its sung form as a recording, what the singer sings, what the chorus sings, what is altered, distorted or masked by the voice synthesizer and the music in the background—all of these are just as important as what exists in the written form. That is why the sung and unwritten aspect of the song becomes as important as the song in its written form and the unwritten aspect of the song is what we call the work of the creative mouth. And because what is sung or spoken is often unclear, not clearly articulated, kept ambiguous, remains 'unsaid', or is made to become one with the music – a musical sound – and so has only a musical or an affective meaning for the listener, the *heard and interpreted aspect* of the song becomes just as important, and this we call the work of the creative ear.

In our time when the song is produced and sold as a recording, the written text is seen as supplementary, and is often treated, like subtitles in a film, as an aid to hearing. Printed versions of songs are seldom officially released by the recording artist. This was not the case when the material and commercial reality of the song was the printed score sheet which also contained the text of the song and from which singers and musicians performed the song. Now, when the recording of the song and text of the song both have material and commercial existence one can notice the differences between the recorded and written forms of the song.

The musical and affective meaning of the song which we see as the work of the creative mouth (whether it's the work of vocal tract of the singer or of the voice synthesizer) and/or the creative ear is not always sensible: it often does not make sense, or sounds strange and weird. It may contain private words or may even be sung in a private language. This sung/heard aspect of the song constitutes the affective and virtual existence of the song and represents the 'becoming-mad' and becoming-creative of the singing voice and the hearing self.

## § 2. Making Sense and Not Making Sense of Musical Non-language Sounds and the Role that Linguistic and Cultural Context Plays in Sense-making

Some singers like Elizabeth Fraser and Jónsi are known for singing songs in unknown and unrecognizable languages or in made-up languages that don't mean anything. Songs like these negate language, and like music, try to express rhythms, moods or affect through pure vocal sound.

Sounds that are expressive only of music place themselves in the domain of music and do not enter the domain of language. But they can be heard and interpreted in linguistically meaningful ways. Musical non-language sounds therefore belong to a category of their own though they are similar to musical language sounds that downplay or manipulate linguistic meaning through vocal effects. Musical non-language sounds are nonverbal vocalizations that are good vehicles for conveying the intensity, ambiguity or lyricality of emotions; they shrink from the extensity, and differentiatedness, of linguistic meaning. These are vocalizations of emotions that want to remain obscure and indistinct, rather than clear and distinct, as clarity and distinctness can rob the emotions of their mystery and of their capacity to evoke moods.

Singing in made-up language is to sing in a language of attitudes and styles; it is the production of the vocal equivalent of facial expressions and body language, and of vocal expressivity which displays a musical personalization or individuation. Singing in non-language sounds – making musical non-lexical vocables – is a pure expression of personalization or individuation.

When I hear the Sigur Rós song 'Hoppipola' written in Icelandic, which I do not understand, and the made-up language, Hopelandic (Vonlenska) the ambiguous sound information forces me to recognize the sounds: and I recognize the alien sound information by projecting onto them what I know and thereby truly hear what only I, as a listener, can hear based on my linguistic and cultural context. As Deleuze and Guattari (2013, p. 362) say, the song contains an act that is similar to the act of skipping, "it jumps from chaos to the beginnings of order in chaos and is in danger of breaking apart at any moment." Below is the transcription of the first few lines based on how I hear it. I am not hearing the song in English; I am just trying to see if I can hear words in the sounds of the song. The underlined lines are words that I recognize.

Kosāmbī

and *uh steer-hreer* key

autumn's *stee* hand *er…*

I can only hear-to-transcribe the first few lines of the song after which the process of hearing-to-transcribe becomes unbearable as I begin to miss the affect-meaning of the song expressed purely through sounds. Sometimes what is moving about a song is just the music-becoming of the vocal sounds and our attempt to understand and hear words in the sounds makes us slip from a state of uncomprehending appreciation, which is a state of intensity, to an undesirable state where we hear words—most of which are nonsensical—but fail to appreciate the intensity and emotion inherent in the sounds.

But the one word that I hear at the beginning of the song, because of the phenomenon of pareidolia, has a meaning that is private and personal: I hear the word Kosambi rather than the Icelandic Brosandi. This is because I don't know Icelandic and am familiar with the word Kosambi which also happens to be a proper name. This is not uncommon as lyrics are often misheard and these misheard words or mishearings in a song are called mondegreens. I can imagine that I heard the word Kosambi because I feel the song represents some kind of awakening and that Kosambi also represents some kind of awakening. But we must not forget that this kind of forming association between heard words and the affective contexts in which they are heard also constitutes sense-making. I associate the heard word with the song by looking for a common factor: both represent a sense of awakening. But I can say my context has informed the way I perceive the song as I heard the word Kosambi and thought the song was about a sense of awakening.

When we say we are affected by the musical sounds of a song what we mean is we are moved unconsciously, which is another way of saying we don't know why we are moved. When we do not know why we are moved we look for reasons and the powerful affect tends to latch onto the one meaningful word that I hear in the song. But we can also say that the word signals the existence of a hidden and repressed world—and this hidden and repressed world, this unconscious, could be as much within us as in the other or *a* collective psyche somewhere in the world. Through the song whose meaning we do not understand but by which we are affected we interact with the unconscious of a world.

But isn't this way of listening to songs a little strange? For one thing, we do not usually transcribe the sounds of a song in order to appreciate it. We also do not transcribe the sounds of a song when we know it is written in a language and for which translation is available. The experience of listening to the song is probably more satisfying when we know what the song means. Here is the English translation of the first stanza of the song:

> Smiling
>
> Spinning round and round
>
> Holding hands
>
> The whole world a blur
>
> But you are standing

This rendering of the sense of the sounds of the song is a little satisfying. But we hear a song for the pleasure of realizing the affect aroused by the power and music of the sounds themselves.

And when we try to make sense of the affect and the sounds of a song by hearing words in them we are probably dealing with 'slips of the ear'. Are there ways in which intellectual and musical affects associate themselves without fully revealing their sense? Not wanting to know the meaning of the song is a way of securing an affective context; we sometimes prevent the flow of sound into predetermined sense in order to secure the ordering of affect inherent in the musical sounds themselves. It is like how we sometimes do not want a definite word mentioned in order to stay with the reality and organization of an indefinite affect. This does not sound so strange if we understand that music is a sensuous medium in which the intensity and musicality of sounds matter more than the sense of the sounds. This way of dealing with musical stimulus, preferring the evocative sensuality of musical expression to the sense of sounds, is a necessary aspect of aesthetic appreciation of music. The problem of the superfluity of sense occurs in music because we tend to regard music as the thing itself.

## § 3. The Kind of Meaning Conveyed by Non-Language Musical Sounds, Our Habitual Ways of Decoding Them and Our Apprehension of Pure Sound

Since we often hear the melody or rhythm of the nonsense syllables and understand them as conveying a certain feeling we habitually employ a thumb-rule to decode the sense of the nonsense syllables or non-language sounds in a song. Often the meaning of the nonsense syllables in a song is the feeling or the mood that they convey and which the meaningful words immediately succeeding the nonsense syllables justify or explain. So the sense of ob-la-di, ob-la-da in the Beatles song of the same name is the meaningful words which immediately succeed it:

Life goes on. The introductory nonsense-syllable melody lal lal la/ lal lal la la la/ lal lal la/ lal lal la la-la—lal lal la/ lal lal la la la/ lal lal la la-la la la la in the Kylie Minogue song carries the lyrical mood which accompanies the situation – conveyed by the words which immediately succeed the sounds – I just can't get you out of my head. Music relies on the lexical codes of language for the realization of semantic meaning.

But do nonsense syllables or musical non-language sounds themselves convey semantic meaning? Bicknell (2002, p. 206) suggests the use of Kant's aesthetic idea to understand the kind of semantic meaning carried by music. Aesthetic idea, according to Kant, is a representation of imagination which brings about much thought but to which no definite thought or concept is commensurate. The form of non-language sounds – as melody, as auditory rhythm, as modulated tone, as non-lexical vocables arising in specific cultures – convey the meaning of affect-saturated experience (a lyrical mood, complication in a relationship, a strange state of being etc.,) which is similar to the meaning of experience conveyed by an aesthetic idea, like an expression in a poem, which is pregnant with many partial representations but whose expression cannot be compassed within a definite concept.

Musical non-language sounds and made-up musical languages are not means for conveying 'semantic meaning'; rather they are immediate ways of languaging the sensorium of experience. Singing in a made-up language is an interesting way of accommodating language as non-language and making music in it to escape the non-musical ordering of affect and thought. It is also a way of making music in a language-like sound material and a way of experiencing music in a language-like sound medium. Non-language sounds are interesting linguistic achievements where vocal sounds are becoming commensurate with the immediacy of sensual and musical experience and the indeterminate richness of aesthetic ideas.

Music is similar to language in the sense that in the tendencies and movements of music, emotion tries to find its own equivalences. Music is different from language in that it is primarily signifying of emotion rather than sense. When a singer sings a melody in nonsense syllables there is a transmission of sound vibrations that is outside language. Similarly, when we listen to a piece of instrumental music we listen to it for its expressive and immersive qualities; our attempt to understand it is often a secondary phenomenon requiring the cognitive mediation of language. But when we are listening to music we are grasping and appreciating relationships that often lie outside the logical relationships that are found in language. There are regions of resonance in the human mind and body that express the Haecceity, the thisness, of the mind and body. The transmission and reception of certain sound-vibrations in the non-language of music is a pure sharing of these resonances of the mind and body.

Expression in the form of sound vibrations is anterior to expression in the form of language. Sound vibrations are also the more immediate and spontaneous carriers of emotions than language because one essentially resonates with sound vibrations. This means sound vibrations have a more direct relationship with the regions of resonance in the human mind and body. Linguistic expression forms, shapes and articulates the resonances so that they begin to exist at the level of understanding. An emotion that we resonate to when also understood gives us the satisfaction of having mastered the emotion and gives birth to intellectual appreciation. This means the mind has made contact with that which moved it and can appreciate it consciously as well as unconsciously.

An unintelligible word in a song that is 'intelligible' only in a rhythmic (or affective) sense would be treated as *Blituri*, the onomatopoeic word used by ancient Greek writers for the sound of the harp- or lyre-string. *Blituri*, for writers of late-antiquity, is a meaningless word, like *tophlattothrattophlottothrat* of Aristophanes. The constructed multilingual onomatopoeia expressing the sound of thunder *bababadalgharaghtakamminarronnkonnbronntonnerronntuonnthunntrovarrhounawnskawntoohoohoord enenthurnuk!* from Joyce's *Finnegan's Wake* is another example of *Blituri.* Deleuze understands *Blituri* as the Stoics did, as a word stripped of meaning which is used with the correlate *Skindapsos*, the machine or instrument. *Blituri* or the blank word when used by the singing voice with vocal sounds that mean something is a sound that is expressive, like the sound of the lyre-string, of pure quality. The ear that apprehends sound as pure quality may or may not choose to make it intelligible but when the sound is expressed as a word it tries to make sense of it. Since making verbal sense of a pure sound word robs it of its quality the listener receives it as, to use a phrase by Diogenes the Babylonian, 'a percussion of the air' which is the proper object of the sense of hearing, or as music made by the instrument of voice.

But when the sound word is transcribed and turned into verbal expression our cognitive system tries to make sense of it. For Deleuze *Blituri* is the blank word whose function is to express the thing. Blank words are usually designated with esoteric words in general such as it and thing. The blank word "says something, but at the same time it says the sense of what it says: it says its own sense." (Deleuze 2005, p. 79) It is, in other words, a paradoxical element, something that is at once word and thing: an emotional truth that can't be explicated.

## § 4. Semantic Localization and How Creative Sense-making of Ambiguous Sound Stimulus Can Discover a Context or Reveal the Unconscious of a Context

The very act of transcribing vocal sound is a form-giving activity which organizes sound matter and utterances into meaningful and sensible forms. When the singer Astrid sings "Your love is like..." followed by the non-language vocal sound, *Eai-ni-naa-naa nun-unn-aaa*, we can employ the thumb-rule to interpret nonsense syllables that we discussed at the beginning of § 3 and interpret the semantic meaning of the non-language sound as "it hurts so good." If we are not satisfied with this meaning the creative ear begins its work of hearing and interpreting: the ear can hear the vocal sound as pure sound which only has affective sense or it can arrange it into affective verbalizations which, when transcribed in a certain way, can carry sense and become part of language. Non-language vocal sounds are different from vocables which are meaningful sounds (like uh-huh, tra la la, dum dee dum, uh-ho) made by people and whose meaning is fixed by their culture and language. In the context of the song "Hurts So Good" the non-language vocal sound is describing what a certain kind of love is like; it can be heard and transcribed as, say:

*Eai-ni-naa-naa nun-unn-aaa*

*Eai-ni-naa-naa nun-unn-aaa*

retaining the expansive rhythm of the vocalization. Or, it can be contracted and abbreviated into a soundword which can then bear a certain sense.

>   *Eaininaanaa Nunnunnaaa*

Sense is added by turning the vocalization into an abstraction and placing the abstraction in the context of sense-making language. So using the lines of the song "Hurts So Good", the creative ear, after hearing the vocal matter as *Eaininaanaa Nunnunnaaa* interprets it as:

>   [When it hurts, but it hurts so good... Can you say it? Can you say it? Your love is like-] *Eaininaanaa Nunnunnaaa*.

The creative work can also look for words in the vocal matter and organize the vocal information into words that are similar to the verbalized vocalization. This is the act of interpreting. One can look for homophonic English words in *Eaininaanaa Nunnunnaaa:*

>   Any *na na*      None None *aaa*

Here, there is a semantic expansion of vocal matter where certain vocal sounds are transformed into words. These sounds are a bit like, but are not, oronyms which are strings of sound that can be perceived as words in two different ways: 'Ice cream'/ 'I scream' and 'Dance early light'/ 'Dawn's early light'. This leads to a further act of creative sense-making in the context of the song:

[Can you say it? Your love is like]

Any *na na*      None None *aaa*

Sense-making here involves giving a language-form to non-language vocal matter, and the language one ascribes to the vocal matter could be one known only to the hearer and not to the singer. When the sense-making of the above vocal matter takes place in a multilingual context of English and Hindi – this is when the context of the sound changes – there is a kind of bracketing of the non-language sound, and it gets a meaning which does not exist in the context of the singer.

Your love is like Any *na na* [no no]

None None *aaa* [come]

Of course, one can further explicate 'None None *aaa*' as 'None, Not one, *aaa* [come]', and make the interpretation appear less strange. But the recognition of familiar words in non-language sounds is indicative of pareidolia or the tendency of perception to impose a recognizable pattern or a meaningful interpretation on a nebulous or ambiguous stimulus. But the imposition of the meaningful interpretation at the linguistic level is assisted by the ability of the non-language sound to facilitate, in the midst of linguistic apprehension, a form of pure pre-linguistic transaction. When non-language sounds are involved sometimes the interaction between a song and a cognitive system results in a kind of affective reaction between the expressive voice and the senses of languages and cultures—and out of this reaction something emerges.

When the sense-making happens in the context of English and Kannada, the rift from the context of the singer is even greater, and it sounds a little strange:

> Your love is like... *Eai ni naan aa*? [Hey, are you me?]
>
> *Naan* one *aaa*? [Am I one?]

This does not usually happen because of a phenomenon that I call *semantic localization*. Semantic localization of sound is a sense-making phenomenon that is parallel to sound localization or the placement of sound in the spatial location of the environment to detect the source of the sound. This plays an important role in how ambiguous auditory stimuli are interpreted. Because of semantic localization we tend to trace the meaning of a sound by going to the source of the sound, that is, we tend to place the ambiguous auditory stimuli in the linguistic and semantic context of the singer or the recording artist.

But semantic localization of ambiguous sound stimuli is only a tendency; it is not always the case that we make sense of ambiguous sound information by going to the source-context of the sound. Instead, we make sense of it by using the context in which the song places it or in which we find the song; the multilingual context of English and Kannada in which the song was interpreted produced a strange result. It revealed something that was hidden and brought out what I wanted to say concerning identity about another song and its music video by the artist called Emotion. This means the ambiguous sound stimulus does work a bit like a Rorschach ink blot: when you begin to interpret or decode it playfully using your immediate context it reveals something hidden in your unconscious. But the important thing to notice here is that the interpretation did not reveal the unconscious of an individual but the unconscious of a context.

But we do not hear or interpret the non-language sounds by turning the sounds into words in a language; rather, we try to find the most appropriate sense for the emotion we feel when we hear the sound in the context of the song. That is why most of the times we hear the sound as making sense in itself and as something that intensifies affect. It is, therefore, hard to say whether the sense of a sound is compressed in the sung sound information. Sense just unfolds in an affective or linguistic context through the act of sense-making. This is what I call the work of the creative ear.

A consciousness in movement is oriented by the stable region of a context which acts as a point of reference which guides and limits what is cognized. Our understanding and experience of context is often complicated by the existence of many points of reference which define a context. But sense-making of the song happens in the context of the relationship between the singer/recording artist and the listener: convention prompts us to assume that the vocal information is produced by the recording artist and is only received by the listener. But the certain way in which the vocal information is heard by the listener depends on the linguistic and affective context and the linguistic and affective resourcefulness of the listener. And these social and individual contexts are productive in their own way since they constitute the extra-bodily hearing and sense-making organs of a cognitive system. I call this extra-bodily sense-making organ of the cognitive system contextual cognitive apparatus or exo-brain.

It is only when we place a song in its larger cultural context, the politico-social context in which it was conceived and performed, that we apprehend its full sense. The sense of the song therefore is not something that can be found exclusively in the printed text, in the sung form, or even in the performed/danced/edited music video. Since the song is something that interacts with existing culture and a people, that is, certain cultural products that have woven the psyche and memory of a collective, the sense of the song is like a comet of singularity that traverses between the singing voice and the written text establishing transversal connections between disconnected experiential lifeworlds and subjective self worlds.

That is why when Kate Bush sings the song 'The Sensual World' inspired by the last part of Molly Bloom's monologue from *Ulysses* it is an event that can't be placed in the singing voice of the singer but an event so much on-the-move, so thrillingly and disorientingly dislocated from the centrality of the self, and so thrown in the elemental frission and excitement of the human lifeworld that we are forced to ask, "Where does man end and culture begin?"

## § 5. The Contextual Cognitive Apparatus or Exo-Brain: How It Is Formed Outside the Brain and Internalized or Interacted with to Enable Cognition

Using non-language sounds in songs is something that musicians do while playing with sound matter. Non-language sounds are often simply verbalizations of rhythm and melody or music made by the instrument of the voice. But sometimes making non-language sounds is a way of expressing mentalese or the form of internal words. Mentalese or the language of thought instead of getting translated into a language gets translated into musical sounds and the affect that these sounds convey. These sounds can be seen as *the unconscious articulations of the unconscious or the articulation of the unconscious as unconscious, that is, as articulations that do not enter the sensible and conscious domain of language-thought.* The singer can therefore say I have expressed something that I do not yet understand or that my voice can express something that my culture and context do not allow me to express in the form of language.

Non-language sounds or what the listener perceives as ambiguous auditory stimuli in a song functions a bit like a Rorschach inkblot but instead of simply drawing something out of our unconscious it enables us to perceive words that are in consonance with our subjective perceptual world or *umwelt.* What is interesting however about ambiguous sound stimulus is that it shrinks away from decoding and explication, refusing to be turned into an extensive magnitude. Rather, by remaining sound and rhythm or by transforming language itself into sound and rhythm it uses the vocal instrument as an instrument of intensification to produce a vibrant vocal matter that is capable of conveying the intensity of affect.

The cognizing mechanism of sensory input – like the auditory and visual input of a filmed song – is in the interpreting schemas of language and culture, and it is when the brain and perceptual mechanism is confronted with ambiguous or unrecognizable sensory data that we realize that the brain looks for available linguistic and cultural resources to decode and interpret unrecognizable sensory information as sensible and intelligible. Without the movement of consciousness or the cognizing mechanism in the extra-brain mechanisms of understanding and interpretation the delayed assimilation of raw experiences and sensory impressions, which continue to exist in us in the form of memory, is not possible. I call the extrabodily cognizing apparatus of sensory information the exo-brain; I also call it contextual cognitive apparatus because we understand sensory information from our immediate frame of reference which is informed by a certain context or by placing the sensory data in a specific context which enables its satisfactory decoding and explication.

Context and location play a very important role in the way artifacts of language grow and develop, and the social context in which artifacts of language adapt and grow and their reception by a people, influences their meaning and significance as much as, or more than, the social context of author who conceived them. So context is not merely something that influences the way we interpret sounds and text, it is the soil in which linguistic life forms grow, develop and bear meaning. A linguistic artifact is interpreted by an entire sensorium – the smell, the flavour, the shade of colour, the quality of light – that it gets from the experiential and cognitive contexts of the individual reader or listener. In fact, artifacts of language, especially those which hide intensity and have an unformed embryonic quality to them, display a form of organic plasticity in relation to the context in which they are placed. This is analogous to the organizer effect, or the principle of induction, discovered by Spemann through tissue transplantation of an embryo during the gastrulation phase. Gastrulation is the stage in embryonic development when "the primordia of the most important organs, the skin and central nervous system, vertebral column and musculature, gut and body cavity" achieve their final dispositions. (Spemann 1935) Parts of the embryo display organic plasticity. (Spemann 1927, p. 180) Experiments conducted by Spemann and Mangold showed that a piece of undifferentiated tissue from the body wall of one embryo in the gastrulation stage implanted onto the body wall of another embryo in the same stage *develops mostly not according to its origins but according to its location*. A presumptive epidermal tissue, that is, a tissue which untransplanted would develop into epidermis, implanted in the area of the brain develops into brain (Uexküll 2010, p. 152) except on a few occasions where it develops according to its origins. And the transplanted portions are interchangeable, that is, the presumptive neural plate could become epidermis and the presumptive epidermis could become the neural plate. An important conclusion that can be derived from this analogy between the plasticity of artifacts of language in relation to their context and the organic plasticity of parts of the embryo at the gastrulation stage is this: psychic processes which lead to or result in the formation of organic artifacts are, or are similar to, vital processes which lead to the formation of organs. Psychical processes which begin to resemble vital processes are affective in nature and the affective intensity and immersion give the process of organic growth of linguistic forms the character of involuntariness and automaticity.

According to the pioneer of neuroplasticity and sensory substitution Bach-y-Rita we do not see with the eyes but with the brain. (Bach-y-Rita et. al. 2005, p. 116) Bach-y-Rita through his tactile-visual sensory substitution device showed that a blind person can experience vision with the help of the artificial eye of the camera mounted on his head which relays visual images, transduced as electrotactile stimulation on the tongue, to the brain. This showed that the human brain is capable of visual perception through the tactile sensory modality of the tongue. The brain, in other words, can see with the camera and the tongue.

Several experiments were conducted at MIT to demonstrate the visual-auditory plasticity of the brain. The eyes of newly born ferrets were wired up to parts of the brain that are normally associated with hearing like the auditory thalamus and auditory cortex, and the cells in the eye which usually made connections with the visual areas of the brain (the visual cortex and the visual thalamus) managed to make connections with, and activate, the auditory parts of the brain, resulting in the ferrets being able to see with parts of the brain associated with hearing. The ferrets did not hear with their eyes but managed to see with parts of the brain not associated with vision. (Sur et.al. 1999; Noë 2009, p .54)

The plasticity of the brain is analogous to the plasticity of the exo-brain, which is the contextual cognitive apparatus of culture which makes us perceive, interpret and experience reality differently, without us losing the certainty that we are perceiving and understanding reality correctly. We can understand a phenomenon or event through a contextual cognitive apparatus not meant to perceive and analyze a certain phenomenon or event and when we do the reality of perception and understanding becomes part of our experiential life world. A foreign-language film with familiar English faces in it dubbed into English will easily be perceived as an English language film; here the contextual cognitive apparatus of English cinema, which is a shared reality, prevents us from perceiving the foreign nature of the film. It is cognitive situatedness again which would make a person from Königsberg perceive Kant as a familiar Russian philosopher who wrote in German while for a German national he would be a proper German philosopher. But it is only when we share the perception and understanding of the subjective life-world of the perceiver that the perception has inter-subjective existence. However, if the experience of the subjective life-world has to have the status of reality there must be social consensus on the right contextual cognitive apparatus that is to be used to perceive, interpret and understand certain phenomena and events.

The same individual A thinks differently about certain issues when he is surrounded by Frenchmen than when he is surrounded by Indians. Formal membership with a group or an organized entity also plays a crucial role in organizing cognition. A as a citizen of France would think differently than A as a citizen of India. This means the set of people who form the milieu of a person and the loyalties that they demand also functions as contextual cognitive apparatus. The greater the degree of transplantation or psychological assimilation with the milieu the greater will be the role of context in aiding a differential cognition.

Contextual cognitive apparatus when internalized becomes the mental context of a person. Mental or cognitive context plays a very important role in how words and symbols are perceived. Two people may read the Sanskrit/Hindi word *shastra* in the same textual context but the mental context of one person makes him read it as *shāstra* meaning 'subject or discipline' while the mental context of the other person makes him read it as *shas-tra* meaning 'weapon'. Mental or cognitive context while enabling the cognition of a certain meaning and pronunciation of a word, and the meaning of a symbol, often filters out meanings and pronunciations which it does not find relevant.

An area called the visual word form area (VWFA) in the left fusiform gyrus is responsive to the visual presence of words and word-like assemblages of letters. Damage to this area of the brain produces deficits in word perception. This visual word form area of the brain which lights up when we are in the visual presence of words has developed because of the cultural habit of reading. In the absence of visual information in the form of words one begins to hear words in the ambiguous sound matter of the song. The existence of a cultural world which has made reading and sense-making mandatory for one's survival makes the existence of non-sense-making expressions and cultural products exotic and interesting. When someone whose neural pathways are developed for word-recognition in a *particular* context or language listens to songs which are sung in unrecognizable languages, or which contain non-language sound words, he perceives words and language of his own context in the ambiguous sound stimuli of the songs. He may also find an emotional milieu for his own intellectual context or find equivalences for his own experience in the experience of the musician as embodied in the song. Such a person would not seek to understand the text of the song but find immersion in a context by engaging with the ambiguous sound matter of the song.

Brain contains engrams which are units of cognitive information imprinted in biophysical and biochemical form. Photographs and videos, as well as pictures like a machine-drawing, a cross-section, an architectural plan with measurements, are engrams outside the brain that induce, influence, and transform memories, so that one remembers and reconstructs the past from the outside rather than from the inside. An archive of family or historical photographs, a set of inscriptions, historical records and documents, which assist in the cognition of the past and the formation of the present and enable sense-making of past and present events can be considered part of the exo-brain or contextual cognitive apparatus. Since the memory of the contextual cognitive apparatus far exceeds the lifespan and memory of the human brain, the brain often contends with the information that could be derived from the outside to reconstruct a time of the past or understand a phenomenon in the present and conceptualizes the contextual cognitive apparatus in a bounded way. Sometimes we use the contextual cognitive apparatus in rather fixed and habituated ways, as a schema, to stabilize the meaning of things. The brain contains a body schema and the body schema exhibits the characteristic of plasticity. It is because of the body schema in the brain that we experience the phenomenon of the phantom limb: the imaginary limb that continues to live even after the loss of the real limb. Once we conceive a contextual cognitive apparatus or exo-brain it produces prototypes of beings that live in and animate a certain context and these prototypes resemble the body schema. The brain harbours prototypes which are psychic manifestations of physical beings and when these prototypes emerge from narratives which are founded on hard reality we call them psychic manifestations of the exo-brain. An extensive body of epic and heroic literature, for example, generates the prototype of the warrior and the mythic type of the ideal man; the body of Romantic literature produces the prototypes of the rebel and the lover. Similarly, histories of modern nations and democracies give birth to the prototypes of the national hero and the citizen. The prototypes which have a general character when they are part of collective unconscious always have a singular character when they are generated by the exo-brain. The death and disappearance of real beings in a certain context does not affect the existence of the prototypes and mythic types as long as they are generated as the schema of a contextual cognitive apparatus. And just as the body schema allows the incorporation of a metal or rubber arm as part of the body, the contextual cognitive apparatus allows the incorporation of entities that can replace or substitute the prototypes.

Thematic and topical connections that are made to understand an object or an entity are connections that exist between ideas, explanations, numbers, images, graphics, texts, historical places, geographical locations, archives, maps etc. The perception of these connections and understanding of entities as part of an interrelated and connected context requires the strengthening of synaptic connections and the formation of large-scale neural circuits that recognize the connected and related nature of disparate materials. Once the neural pathways become sufficiently strong through learning and repeated exposure to a certain topic in its interrelated nature thought automatically flows and begins to inhabit a context. This inhabiting of thought in a connected network of entities happens because entities in the world are connected together outside the brain through associations and discursive structures that bring into being a context. The very act of traversing an area of learning experientially and acting in it allows the brain to make synaptic connections and grow neural pathways and form neural circuits that enables thought to inhabit a context. This contextual apparatus which exists outside the brain performs the function of the brain by enabling cognition and interpretation of discrete entities and by allowing thought to inhabit and flow. This is why we call the contextual cognitive apparatus the exo-brain.

Cognition stabilizes itself to a context; it begins to treat a well-defined context as a stream of space-time which has its own reality. At the most fundamental level context is not outside the brain but is the frame that we impose on a diffuse reality to define and structure it in a certain way. For example, while travelling from place A to place B if I see the road as the route to be traversed to reach place B, I perceive the road and experience riding the motorbike in a certain way. But if I conceive the road as a racetrack I perceive the road and the fellow riders on the road in an entirely new way, it changes what I see on the road and the way I ride the motorbike: now I see the entire constellation of the road as a track that I must navigate by overcoming the obstacles of other vehicles—my focus being only to outrace the other drivers on the road. And even if there is one other person on the road racing with me with the intention of reaching place B, the intersubjective reality of the race is created. Thus cognition not only stabilizes itself to a context, that is not only does it alter, and structure, our perception to carve out *a* context from the disparate sense data, but gaining foothold in external reality, it modulates our experience of reality.

Consciousness is about sense-making, the becoming-expansive of willing. When we consider consciousness for understanding we discern in its continuity discrete sensory and semantic wholes which are made up of perceptual units and/or memory units and a general sense of how they make us feel and what they signify——this is what makes up our consciousness. When we remember something we not only remember a sensory impression but also what the sensory impression signified to us in terms of how we felt and when we had that impression and what it meant to us. Sensory impressions or pictures of the world always associate themselves with particular affects and particular significations. At higher levels of consciousness (where we are trying to make sense) we do not act on an impulse but consider the options we have in acting or in not acting. Consciousness is the aware space where we process the sensory input and actively will the output through our actions. In proposing the concept of the contextual cognitive apparatus or exo-brain I am stating that the form of being conscious in terms of perceiving the input, processing it, and willing the output happens because of explanatory and knowledge mechanisms outside the brain which enable and influence the cognition of sensory input in certain specific ways, and which also allows for a rigorous processing, or understanding, of the input, so that it gains in meaning and leads to a considered response. The input that the brain receives is mediated by cognitive mechanisms in which we are situated but is also processed with its help so that we make judgments and decisions that are meaningful and impactful. Cognitive processing of input then is not something that happens only in the brain but something that happens in our interaction with the environment, and this interaction with the environment in the human world, is mediated through contextual cognitive apparatus or exo-brain. It is only when we perceive ambiguous sound stimuli as meaningful either in terms of it having an affective meaning or in terms of it having a sensible meaning that the contextual cognitive apparatus or exo-brain operates at the level of perception and at the level of sense-making. The assumption that the perceived stimulus is meaningful and valuable transforms the very nature of perception from one that is instantaneous and immediate to one that is extended, engaged and interactive—requiring a *transaction* with the world.

## § 6. Contextual Cognitive Apparatus or Exo-Brain Can Proceed from an Encounter with the Other as the Encoding of the Interactions of an Intelligence with that Event

How does an encounter with the other—with another culture, with another set of values, with another way of relating to the world, with another mythological universe, with another context—change us? If it disturbs us how do we cope with the disturbance? It is possible that the encounter initiates a dialectical process of inter-cultural or context-sensitive dialogue which leads to an act of synthesis. We look for commonalities that make the other familiar and similar to our own culture. We also develop an irritable attitude towards the noisy aspects of our own culture, and reject them from our personal culture. This is the beginning of the adoption of a context, and the context is adopted when we come in contact with a contextual cognitive apparatus which has been formed as a result of an encounter with the other.

An encounter with another culture brings about a change in one's personal culture, creating a rift between personal culture and public culture. If the personal culture is a living culture, if the tradition one has imbibed is part of the living tradition, one feels at home in one's personal culture. Otherwise the imbibed personal culture is externalized as archival memory or cultural memory in different forms of recordings like written literature, scientific and artistic models etc., which capture the unfolding of an event or the organization of organized activities. One consciously records the unfolding of an event or the organization of an organized activity by staging it and by thinking through the different stages of its change and development. And the turning of subjective experience into objective consciousness of individuals and collectives with causal properties transforms the result of the encounter with the other into a recording of the event outside the brain. The exo-brain, when it closely matches subjective experience, therefore has a consciousness that is peculiar to the consciousness which results from an encounter with the other, it plays out objectively and impersonally almost like a repetition-with-a-difference of an event in a consciousness, like a simulation of forces that the brain enters into in order to relive the ramifications of an event. The interaction of an agent with the context in which an event unfolds is the intelligence that is encoded in a contextual cognitive apparatus or exo-brain. Therefore, with contextual cognitive apparatuses we understand entities through events that transformed them and morphed them into their current form of life.

Most things that exist (there are some things whose existence we can only infer based on the effect that they have on things that are accessible) exist in a way that it is accessible by a consciousness or had existed in a consciousness in a way that it is available to us. Exo-brain when it is in interaction with a brain has the consciousness of the existence of things, events and ideas outside the brain in a way that it is available to, and can reorganize and situate, a brain. The consciousness of the exo-brain is something that one inherits by interacting with the exo-brain and with other brains which interact with it.

## Discography